# ACTIVE ILLUMINATION USING A DIGITAL MICROMIRROR DEVICE FOR QUANTITATIVE PHASE IMAGING

SEUNGWOO SHIN,[1,2] KYOOHYUN KIM,[1,2] JONGHEE YOON,[1] YONGKEUN PARK[1,*]

[1]Department of Physics, Korea Advanced Institute of Science and Technology, Daejeon 305-701, Republic of Korea
[2] Contributed equally to this work
*Corresponding author: yk.park@kaist.ac.kr

**We present a powerful and cost-effective method for active illumination using a digital micromirror device (DMD) for quantitative phase imaging techniques. Displaying binary illumination patterns on a DMD with appropriate spatial filtering, plane waves with various illumination angles are generated and impinged onto a sample. Complex optical fields of the sample obtained with various incident angles are then measured via Mach-Zehnder interferometry, from which a high-resolution two-dimensional synthetic aperture phase image and a three-dimensional refractive index tomogram of the sample are reconstructed. We demonstrate the fast and stable illumination control capability of the proposed method by imaging colloidal spheres and biological cells, including a human red blood cell and a HeLa cell.**

Quantitative phase imaging (QPI) has emerged as an invaluable tool for imaging small transparent objects, such as biological cells and tissues [1, 2]. QPI employs various interferometric microscopy techniques, including quantitative phase microscopy [2] and digital holographic microscopy [3], to quantitatively measure the optical phase delay of samples. In particular, the measured optical phase delay provides information about the morphological and biochemical properties of biological samples at the single-cell level. Recently, QPI techniques have been widely applied to study the pathophysiology of various biological cells and tissues, including red blood cells (RBCs) [4-7], white blood cells [8], bacteria [9-11], neurons [12-14], and cancer cells [15, 16].

Controlling the illumination beam is crucial in QPI. Especially for measuring 3-D refractive index (RI) tomograms [17] or 2-D high-resolution synthetic aperture images [18], angles of plane wave illumination impinging onto a sample should be systematically controlled, and the corresponding light field images of the sample should be measured. Traditionally, galvanometer-based rotating mirrors have been used to control the angle of the illumination beam. A galvanometer-based rotating mirror located at the plane that is conjugate to a sample can control the angle of the incident beam by tilting the mirror with a certain electric voltage.

The use of galvanometers, however, has several disadvantages. Inherently, there exists mechanical instability due to position jittering induced by electric noise and positioning error at high voltages. When a two-axis galvanometer is used, the rotational surfaces of each axis cannot be simultaneously conjugate to a sample due to its geometry, and this may induce unwanted additional quadratic phase distribution on the illumination beam that can limit the accurate measurements of 3-D RI tomograms. To solve this optical misalignment, two one-axis galvanometers can be placed at separate conjugate planes, but it requires a bulky optical setup with a long optical path, which can deteriorate phase noise. More importantly, galvanometers cannot generate acomplex wavefront; only tilting of a plane wave is permitted. Recently, a spatial light modulator (SLM) was used to control the angle of illumination by displaying a linear phase ramp on the SLM which is located at the plane that is conjugate to the sample [19]. Because an SLM does not contain mechanically moving parts, the measurements of 3-D RI distribution were performed in high stability. However, the intrinsically slow response of liquid crystal realignment in SLMs significantly limits acquisition speed, and the high costs of an SLM prevent wide application.

In this letter, we propose a novel active illumination control method that uses a digital micromirror device (DMD) for QPI techniques. A DMD consists of hundreds of thousands switchable micromirrors, which can be individually controlled in on-and-off states. Displaying binary hologram patterns on a DMD with appropriate spatial filtering, plane waves with various illumination angles are generated and impinged onto a sample. Then, complex optical fields of the sample obtained with various incident angles are measured via Mach-Zehnder interferometry for 3-D RI tomography or 2-D high-resolution synthetic aperture imaging. Here, we validate the use of the DMD for QPI techniques by experimentally demonstrating the 3-D RI tomography or 2-D high-resolution synthetic aperture imaging of individual poly(methyl methacrylate) (PMMA) beads, human RBCs, and HeLa cells.

The use of a DMD has several advantages for quantitative phase microscopy and digital holographic microscopy,due to the ultra-high modulation speed reaching up to a few kHz, high mechanical stability, and the ability to generate arbitrary wavefronts. Furthermore, the use of the DMD is cost effective and easily compatible with existing QPI instruments.

The experimental setup is shown in Fig. 1. The setup consists of a Mach-Zehnder interferometric microscope combined with an illumination control part that uses a DMD. A laser beam ($\lambda$ = 532 nm, SDL-532-100T, Shanghai Dream Laser Co., China) is split into two arms by a 2×2 single-mode fiber optics coupler (FC532-50B-FC, Thorlabs, NJ, USA). One arm is used as a reference beam, and the other arm, which is used for a sample beam, is reflected by a DMD (DLP® LightCrafter™ 3000, Texas Instruments Inc., TX, USA) before illuminating a sample. The DMD is located at the conjugate plane of the sample plane, andit displays a binary pattern, the Lee hologram, to control the wavefront of the reflected beam [20, 21]. The detailed principle of the DMD control for the generation of plane waves with desired the illumination angles is described in the following paragraph..

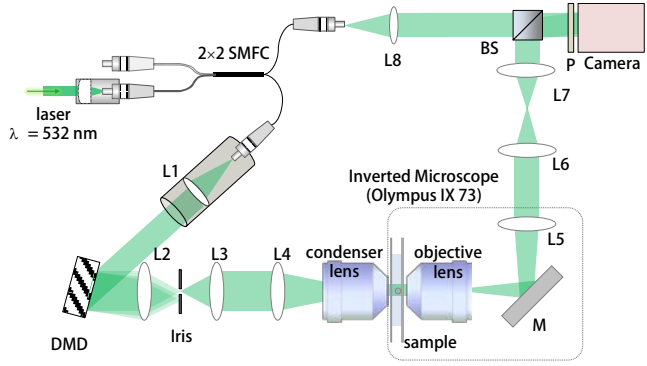

Fig. 1. Experimental setup. The laser beam is divided by a 2×2 single-mode fiber optics coupler (2×2 SMFC) into a sample and a reference arm. The illumination angle of plane waves impinging on a sample is controlled by a digital micromirror device (DMD) and the iris. L1–8: lenses ($f_1 = 25$ mm; $f_{2-4} = 250$ mm; $f_5 = 180$ mm; $f_{6-7} = 200$ mm; $f_8 = 75$ mm); BS: beam splitter; P: linear polarizer.

The beam reflected by the DMD is diffracted in several diffraction orders, among which the 1st-order diffracted beam is selected by using an iris located at the Fourier plane of the DMD plane. The angle-scanning range of the 1st-order diffraction was 1° at the plane of the DMD due to the size of the micromirrors. The angle-scanning range is then magnified and spans the numerical aperture (NA) of a condenser lens, using a tube lens (L4, $f = 250$ mm) and a condenser lens (UPLSAPO 60XW, 60×, NA = 1.2, Olympus Inc. Japan).

Then, a plane wave with a desired illumination angle impinges on a sample on the stage of an inverted microscope (IX 73, Olympus Inc.), and the diffracted beam from the sample is collected by an objective lens (UPLSAPO 60XW, 60×, NA = 1.2, Olympus Inc.) and a tube lens (L5, $f = 180$ mm). The sample beam is further magnified by an additional 4-$f$ telescopic imaging system with lenses L6–7. The sample beam is projected onto a charge-coupled device (CCD) camera (Lt365R, Lumenera Co., Canada), where it interferes with a reference beam with a slight tilt controlled by a beam splitter to generate a spatially modulate off-axis hologram. From the measured hologram, the optical field image of the sample, containing the amplitude and phase information, is retrieved via phase retrieval algorithms [22].

A DMD consists of hundreds of thousands switchable micromirrors, which can be individually controlled in on-and-off states with ultra-high speed reaching up to a few kHz. The DMD was originally designed for a rapid intensity controller for a digital light-processing (DLP) projector [23]. To achieve phase-only control of the illumination wavefront using a DMD, we exploited binary amplitude off-axis holography [21, 24].

To control the illumination angle using the DMD, we first generated the binary amplitude off-axis holograms, or Lee hologram[24]. An ideal amplitude hologram, $f(x, y)$, having a desired phase map, $\phi(x, y)$, can be described as follows:

$$f(x,y) = \frac{1}{2}\left[1+\cos\left\{2\pi ux + 2\pi vy + \phi(x,y)\right\}\right] \quad (1)$$
$$= \frac{1}{2} + \frac{1}{4}\exp\left[j2\pi(ux+vy)\right]\exp\left[j\phi(x,y)\right] + \frac{1}{4}\exp\left[-j2\pi(ux+vy)\right]\exp\left[-j\phi(x,y)\right],$$

where $u$ and $v$ are carrier frequencies along the spatial coordinates $x$ and $y$, respectively. Because a tilt in a plane wave is described as a linear phase ramp, the desired wavefront has a phase map as $\phi(x, y) = 2\pi (u'x + v'y)$, where $u' = \cos\theta_x/\lambda$ and $v' = \cos\theta_y/\lambda$ represent the spatial frequencies of the tilted plane wave with tiling angles $\theta_x$ and $\theta_y$ along the spatial coordinates $x$ and $y$, respectively. The binary amplitude hologram, $h(x, y)$ is obtained by thresholding $f(x, y)$, i.e. $h(x, y) = 1$ for $f(x, y) > 1/2$, and $h(x, y) = 0$ otherwise.

When the binary amplitude hologram, $h(x, y)$ is displayed on the DMD, reflected light from the DMD contain several orders of diffraction. Among them, the first term in Eq. (1) corresponds to the unmodulated zero-th order beam; the second and the third term are the 1st- and -1st-rder diffracted beams, respectively [Fig. 2(a)]. Because the desired wavefront is the 1st diffraction beam, an aperture stop (the iris in Fig. 1) is located at the Fourier plane of the DMD to block the other diffraction beams. When the aperture stop is aligned along the 1st diffraction, the carrier frequencies $u$ and $v$ become zero in this new optic axis; thus, the resultant illumination has a phase ramp of $\phi(x, y) = 2\pi (u'x + v'y)$.

To demonstrate the validity of the use of the DMD for controlling the angle of an incident beam, we first measured a 2-D synthetic aperture phase image and the 3-D RI distribution of a PMMA microsphere with a diameter of 3 μm ($n = 1.4934$ at $\lambda = 532$ nm, Sigma-Aldrich Inc., MO, USA) immersed in index-matching oil ($n = 1.4527$ at $\lambda = 532$ nm, Cargille Laboratories, NJ, USA).

To effectively illuminate spanning the NA of the condenser lens while minimizing scanning time, we scanned the illumination angle in a circular pattern in addition to the normal illumination. The representative Lee holograms displayed on the DMD are shown in Fig. 2(b) with the desired spatial frequencies $(u', v')$. After the spatial filtering by the irise, the plane waves with corresponding incident angles illuminate a sample. The holograms of the sample obtained with plane waves with corresponding illumination angles are shown in Fig. 2(c). The optical field images were retrieved from the corresponding holograms (results not shown). The 2-D Fourier spectra of corresponding optical field images are shown in Fig. 2(d), which shows the scanning of the high intensity peak corresponding to the undiffracted illumination beams. This clearly demonstrates the systematic control of illumination beams using the DMD.

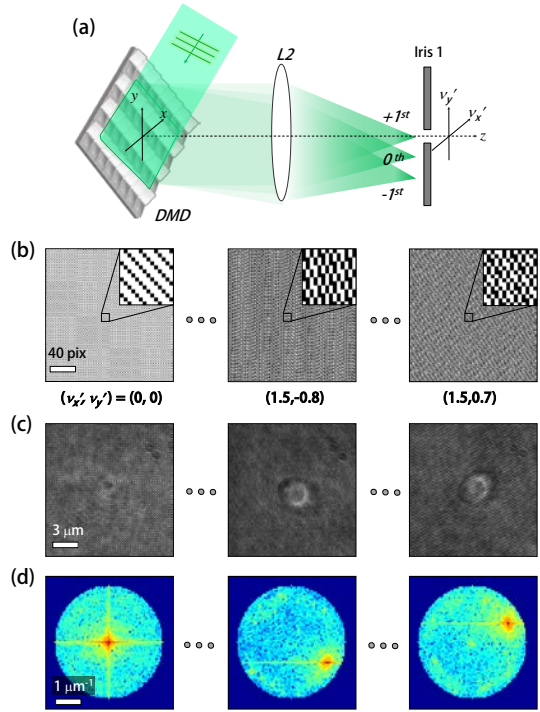

Fig. 2. Illumination angle control using a digital micromirror device (DMD) (a) Lee holograms are displayed on the DMD, and the 1st diffraction beam containing the wanted wavefront is filtered out by iris. (b) A series of binary amplitude off-axis holograms displayed on the DMD. *Inset*, magnified patterns. (c) Holograms of a sample, poly(methyl methacrylate) bead, obtained with illumination controlled by the DMD. (d) Corresponding Fourier spectra of the sample.

From the multiple 2-D holograms obtained with various controlled angles of illumination, a 2-D synthetic aperture image and a 3-D RI tomogram of the microsphere were reconstructed (Fig. 3). The 2-D synthetic aperture image was coherently synthetized from the measured multiple 2-D holograms with various illumination angles [18], resulting in increased spatial resolution and signal-to-noise ratio (SNR) [25]. The 3-D RI tomogram was constructed using an optical diffraction tomography algorithm [17, 26-28].

The generated 2-D synthetic aperture phase image shows a high-resolution phase image of the bead with an enhanced signal-to-noise SNR [Fig. 3(a)]. The reconstructed 3-D RI tomogram shows that the morphology and the retrieved RI value of the bead are in good agreement with the manufacturer's specification [Fig. 3(b)].

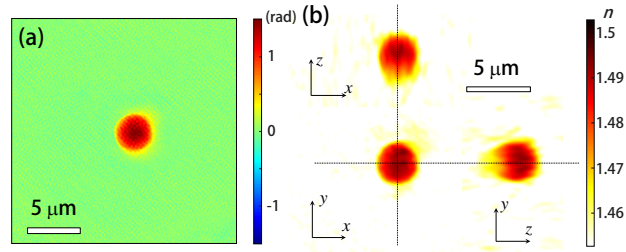

Fig. 3. (a) High-resolution 2-D synthetic aperture phase image and (b) cross-sectional slices of a reconstructed tomogram in *x-y*, *x-z*, and *y-z* planes of a PMMA bead with the diameter of 3 μm.

To further demonstrate the applicability of the present method, we measured the 3-D RI distributions of individual human RBCs and HeLa cells. RBCs were obtained from a healthy volunteer and were prepared according to the standard sample preparation protocol [29]. In brief, the blood was diluted 300 times in Dulbecco's phosphate buffered saline (PBS) solution (Welgene Inc., Republic of Korea), and sandwiched between coverslips before measurements. HeLa cells were maintained in Dulbecco's Modified Eagle's Medium (DMEM, Welgene Inc., Republic of Korea) supplemented with 10% heat-inactivated fetal bovine serum, 4,500 mg/L D-glucose, L-glutamine, 110 mg/L sodium pyruvate, sodium bicarbonate, 1,000 U/L penicillin, and 100 μg/mL streptomycin. Cells were subcultivated on a 24×40 mm cover glass (Marienfeld-Superior, Germany) for 4 – 8 hours before the experiments, and then they were washed with PBS solution immediately before measurements. Next, 400 μl of prewarmed DMEM was added to the sample, and then another cover glass was sandwich the sample to prevent the sample drying.

The 3-D RI distribution of a human RBC was measured and reconstructed. Cross-sectional slices of the 3-D RI tomogram in the *y-z* and *x-z* planes are shown in Fig. 4(a), which clearly show the characteristic biconcave shape of healthy RBCs. The reconstructed 3-D RI distribution of a HeLa cell shows complex cell morphology [Fig. 4(b)].

The cross-sectional slice of the tomogram of the RBC in the *x-y* plane presents the homogenous distribution of the RI value, which implies the uniform distribution of hemoglobin (Hb) protein in RBC cytoplasm. The Hb concentration of RBCs is directly related to the RI value because RBCs mainly consist of Hb solution. There is a linear relationship between the concentration and the RI of Hb solution [30], as $C = (n - n_m)/\alpha$, where $C$ is the Hb concentration, $\alpha$ is the RI increment ($dn/dc = 0.148$ ml/g at $\lambda = 532$ nm for oxygenated Hb)[31, 32], and $n_m$ is the RI value of the PBS solution ($n_m = 1.337$ at $\lambda = 0.532$ nm).

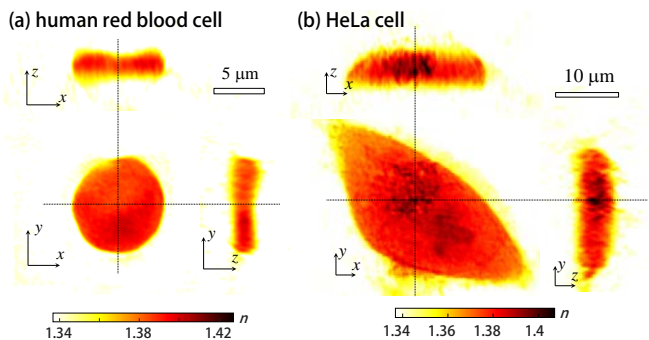

Fig. 4. Cross-sectional slices of reconstructed tomograms in *x-y*, *x-z*, and *y-z* planes of (a) a human red blood cell and (b) a HeLa cell, respectively.

The average value of the measured RI is $1.388 \pm 0.002$, which corresponds to the Hb concentration of $34.47\pm1.42$ g/dL. Furthermore, the cell volume and the total Hb contents of an RBC can also be obtained from the measured RI tomogram. The cell volume was calculated as $88.18\pm3.87$ fL by counting the number of voxels having RI values larger than the RI threshold, defined as 50% of the maximum RI contrast of RBCs in PBS solution [33]. The total amount of Hb contents of individual RBCs was calculated as $30.35 \pm 0.59$ pg from the summation of Hb concentration within the volume of RBCs. All calculated parameters are within the normal range of the healthy physiological condition of RBCs.

In summary, this letter presents the use of a DMD for the control of beam illumination for QPI techniques. Displaying a series of Lee holograms on the DMD with appropriate spatial filtering, the angle of the beam illumination on a sample is systemically controlled with high speed and high stability. With controlled illumination using the DMD, we demonstrated synthetic aperture phase imaging and optical diffraction tomography by measuring the optical fields of samples including individual colloidal spheres and biological samples, including human RBCs and HeLa cells.

Although demonstrated by Mach-Zehnder interferometry in this letter, the proposed illumination approach is sufficiently broad and general that it should directly find application in other quantitative phase microscopy or digital holographic microscopy techniques. For example, by introducing a DMD to the illumination part, an existing bright-field microscope can be converted into a 3-D quantitative phase microscope utilizing the quantitative phase imaging unit [34, 35]. Furthermore, this method can also be combined with other QPI modalities, such as spectroscopic [36-39] and polarization-dependent [40, 41] phase imaging with various incident angles because the mirror array on the DMD does not suffer from dispersion or birefringence, while liquid-crystal-based SLMs do.

One of the advantages in the use of DMDs for QPI is their ultra-fast display rate. Commercially available DMDs have a full-frame display rate up to 32 kHZ. When combined with a high-speed camera, the maximum acquisition rate for 3-D RI tomograms can reach up to 3 kHz because 10 different illumination beams are enough to construct a robust tomograms [42, 43].

It is also noteworthy that the present approach can generate complex wavefront illumination for QPI. For example, various spatial frequencies are multiplexed to generate different spatially modulated optical properties, such as polarization [44]. The degree of spatial coherency of illumination beams can also be controlled by multiplexing spatial frequencies; as an extreme example, speckle illuminations [45] can be realized by the present approach. In addition, the cost of a DMD is approximately 100 USD, and the implementation of a DMD in an existing microscopic setup is straightforward. Thus, we envision that DMDs will be widely used for active illumination schemes for quantitative phase microscopy and digital holographic microscopy.

**Funding.** KAIST, KAIST-Khalifa University Project, APCTP, the Korean Ministry of Education, Science and Technology (MEST), and the National Research Foundation (2012R1A1A1009082, 2014K1A3A1A09063027, 2013M3C1A3063046, 2012-M3C1A1-048860, 2014M3C1A3052537).

**Acknowledgment**. We thank Mr. Hyeonseung Yu for helpful comments about the DMD control.